\documentclass[structabstract]{aa}  
\usepackage{times}
\usepackage{natbib}
\usepackage{cite}
\usepackage{float}
\usepackage{graphicx}
\usepackage{longtable}
\usepackage{supertabular}
\usepackage{lscape}
\usepackage{txfonts}
\newcommand{\msun}{\ifmmode M_{\odot} \else $M_{\odot}$\fi}
\newcommand{\rsun}{\ifmmode R_{\odot} \else $R_{\odot}$\fi}
\newcommand{\kms}{$km\,s$^{-1}}

\begin{document}
   \title{Plaskett's Star: Analysis of the CoRoT photometric data \thanks{The CoRoT space mission was developed and is operated by the French space agency CNES, with participation of ESA's RSSD and Science Programmes, Austria, Belgium, Brazil, Germany and Spain.}}


     \author{L. Mahy \inst{1} \and E. Gosset \inst{1}\fnmsep\thanks{Senior Research Associate F.R.S.-FNRS} \and F. Baudin \inst{2} \and  G. Rauw \inst{1}\fnmsep\thanks{Research Associate F.R.S.-FNRS} \and M. Godart \inst{1} \and T. Morel \inst{1} \and P. Degroote \inst{3} \and C. Aerts \inst{3,5} \and R. Blomme \inst{4} \and J. Cuypers \inst{4} \and A. Noels \inst{1} \and E. Michel \inst{6} \and A. Baglin \inst{7} \and M. Auvergne \inst{8} \and C. Catala \inst{8} \and R. Samadi \inst{8}}

   \institute{Institut d'Astrophysique et de G\'eophysique, University of Li\`ege,
              B\^at. B5C, All\'ee du 6 Ao\^ut 17, 4000 Li\`ege, Belgium\\
              \email{mahy@astro.ulg.ac.be}
              \and 
              Institut d'Astrophysique Spatiale (IAS), B\^atiment 121, 91405, Orsay Cedex, France
              \and
              Instituut voor Sterrenkunde, K.U. Leuven, Celestijnenlaan 200D, 3001 Leuven, Belgium
	      \and
	      Royal Observatory of Belgium, Ringlaan 3, 1180 Brussel, Belgium
              \and
              Department of Astrophysics, IMAPP, University of Nijmegen, PO Box 9010, 6500 GL Nijmegen, The Netherlands
              \and
              LESIA, CNRS, Universit\'e Pierre et Marie Curie, Universit\'e Denis Diderot, Observatoire de Paris, 92195 Meudon Cedex, France
              \and
              Laboratoire AIM, CEA/DSM-CNRS-Universit\'e Paris Diderot; CEA, IRFU, SAp, centre de Saclay, 91191, Gif-sur-Yvette, France
              \and
              LESIA, UMRR8109, Universit\'e Pierre et Marie Curie, Universit\'e Denis Diderot, Observatoire de Paris, 92195 Meudon Cedex, France
   }


 
  \abstract
   {The second short run (SRa02) of the CoRoT space mission for Asteroseismology was partly devoted to stars belonging to the Mon\,OB2 association. An intense monitoring has been performed on Plaskett's Star (HD\,47129) and the unprecedented quality of the light curve allows us to shed new light on this very massive, non-eclipsing binary system.}
   {We particularly aimed at detecting periodic variability which might be associated with pulsations or interactions between both components. We also searched for variations related to the orbital cycle which could help to constrain the inclination and the morphology of the binary system.}
   {An iterative Fourier-based prewhitening and a multiperiodic fitting procedure have been applied to analyse the time series and extract the frequencies of variations from the CoRoT light curve. We describe the noise properties to tentatively define an appropriate significance criterion and, in consequence, to only point out the peaks at a certain significance level. We also detect the variations related to the orbital motion and study them by using the \texttt{NIGHTFALL} program.}
   {The periodogram computed from Plaskett's Star CoRoT light curve mainly exhibits a majority of peaks at low frequencies. Among these peaks, we highlight a list of about 43 values, including notably two different sets of harmonic frequencies whose fundamental peaks are located at about $0.07$ and $0.82$~d$^{-1}$. The former represents the orbital frequency of the binary system whilst the latter could probably be associated with non-radial pulsations. The study of the $0.07$~d$^{-1}$ variations reveals the presence of a hot spot most probably situated on the primary star and facing the secondary.}
   {The investigation of this unique dataset constitutes a further step in the understanding of Plaskett's Star. These results provide a first basis for future seismic modelling and put forward the probable existence of non-radial pulsations in Plaskett's Star. Moreover, the fit of the orbital variations confirms the problem, already mentioned in previous works, of the distance of this system. The existence of a hot region between both components renders the determination of the inclination ambiguous.}


   \maketitle
%

\section{Introduction}

Plaskett's Star, or \object{HD\,47129}, has long been considered as one of the most massive binary systems in our Galaxy. For nine decades, this star has not stopped fascinating by its complexity. It has been the subject of several investigations in different wavelength domains but, despite these efforts, the system is still not fully understood. 

Plaskett's Star is a non-eclipsing binary \citep{mor78} composed of two very massive and luminous O-type stars. The secondary component features broad and shallow stellar lines suggesting that this star rotates rapidly. Its projected rotational velocity has been estimated close to $300~\kms$ while that of the primary has been measured at about $75~\kms$ \citep{lin08}. As a consequence of this high rotation speed, the secondary star probably presents a temperature gradient between the poles and equator which could bias the determination of its spectral type but, most importantly, the secondary has a rotationally flattened wind. This configuration probably accounts for several properties of the wind interaction zone. The study of the H$\alpha$ region by \citet{wig92} and \citet{lin08} as well as the analysis of \citet{lin06} in the X-ray domain have confirmed this assumption.

\citet{Bag92} applied a tomographic technique to the International Ultraviolet Explorer (IUE) data to separate the contribution of the secondary star from the primary spectrum. Their investigation of the individual spectral components has provided the spectral types of O7.5~I and O6~I for the primary and secondary stars, respectively, and a mass ratio of $M_2/M_1 = 1.18 \pm 0.12$. Assuming an inclination of $71 \pm 9^{\circ}$ as estimated by \citet{Rud78} from polarimetry, they found masses equal to $42.5~\msun$ for the primary and $51.6~\msun$ for the secondary.

\citet{lin08} found, from high-resolution optical spectra, minimum masses of $45.4$ and $47.3~\msun$ for the primary and secondary, respectively, implying a mass ratio of about $1.05 \pm 0.05$. They also disentangled the spectra by using an algorithm based on the method of \citet{gl06}. The individual spectra indicated an O8~III/I$+$O7.5~V/III binary system. In addition, these authors used the model atmosphere code CMFGEN \citep{hm98} to derive the wind and the photospheric properties of both components. The major point of their study is the confirmation of a N overabundance and a C depletion of the primary star, providing additional proof for a binary system in a post Roche lobe overflow evolutionary stage where matter has been transferred from the primary to the secondary star.

The most disturbing point concerning Plaskett's Star is however the discrepancy between the luminosity of both components and their dynamical masses. \citet{lin08} explained that these stars have spectral types which are too late for their masses. A solution to this problem is to assume a larger distance of the star although this would imply that Plaskett's Star does not belong to the \object{Mon~OB2} association.
Despite the numerous investigations undertaken to probe the physics of this binary system and its components, Plaskett's Star still hides a part of mystery. 

Even though \citet{Ree09} have emphasized the difficulty to detect pulsation modes in rapidly rotating stars, there are some examples of rapidly rotating O~stars ($\zeta$~Oph, \citealt{kam97}; HD\,93521 \citealt{how93}, \citealt{rau08}) where spectroscopic and photometric variability, probably related to non-radial pulsation modes with periods of a few hours, have been identified. These rapid rotators have properties quite reminiscent of those of the secondary component in Plaskett's Star, which also displays line profile variability \citep{lin08} although the existing spectroscopic data of the system are too scarce to characterize properly these variations. Asteroseismology could therefore provide new insight into the properties of the components of Plaskett's Star.

For this purpose but also to further constrain the binary system itself, Plaskett's Star has been chosen as one of the CoRoT \citep[Convection, Rotation and planetary Transits,][]{bag06,auv09} satellite targets in the Asteroseismology field. The unprecedented quality of the CoRoT light curve clearly allows us to search for variations linked to the orbital period of the system and to determine the possible existence of very low-amplitude variations due to, e.g., the presence of radial and non-radial pulsation modes.

The present paper describes a complete and detailed analysis of the light curve of Plaskett's Star observed by the CoRoT satellite. We organize it as follows. In Section~\ref{corot_ana}, we present the CoRoT data. Section~\ref{corot_result} is devoted to the frequency analysis and to a study of the noise properties of the data. In Section~\ref{corot_disc}, we discuss the inclination of the binary system and some other results obtained from the analysis of the light curve. Finally, we set forth our conclusions in Section~\ref{corot_concl}.

\section{The CoRoT data}\label{corot_ana}

Plaskett's Star was observed during the second CoRoT short run SRa02 pointing towards the anticenter of the Galaxy. The observations were made between October 08, 2008 ($HJD = 2\,454\,748.485467$) and November 12, 2008 ($HJD = 2\,454\,782.819956$). Hereafter, we will express the date as $HJD-2\,450\,000$. The time series obtained is spread over $\Delta T = 34.334489$ days, with a sampling of one point every $32$~s. This involves a frequency resolution of $1/\Delta T = 0.029125$~d$^{-1}$.

\begin{figure}[htbp]
\centering
\includegraphics[width=9cm, bb=22 260 564 679,clip]{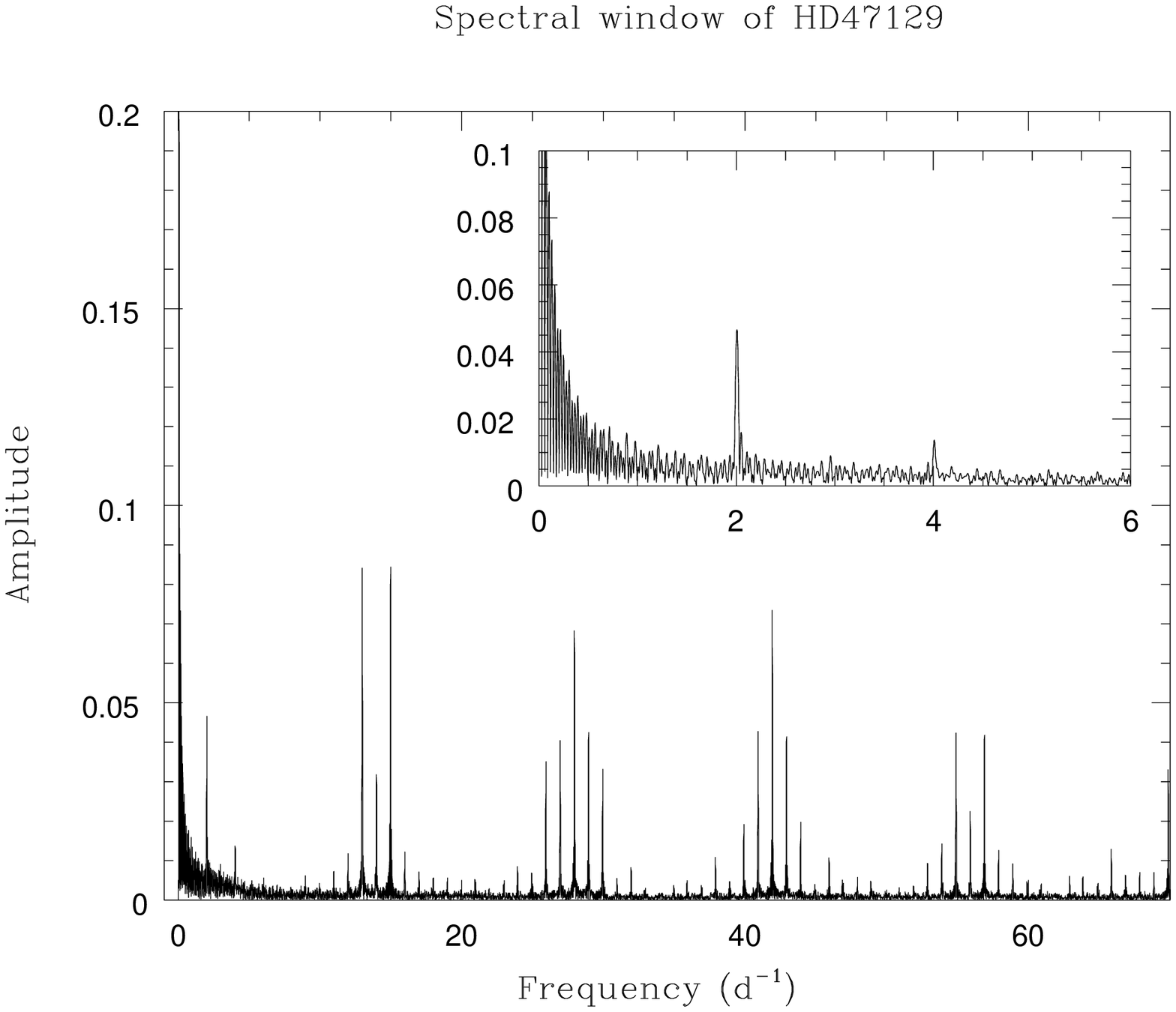}
\caption{Spectral window of the final version of Plaskett's Star light curve, observed during the second short run of the mission. \label{spectralwin}}
\end{figure}

The raw light curve of Plaskett's Star contains 92696 points. At first, we discarded flagged points potentially corrupted by the instrumental conditions (e.g., the changes of CCD masks or the generation of outliers). These observations represent about 4.2\% of the datapoints of the CoRoT light curve of Plaskett's Star. In addition, we searched the CoRoT light curve for possible jumps or the presence of discontinuities due to a change of the CCD temperature and we corrected all of them. Secondly, we also discarded the flagged points associated with the environmental perturbation of the satellite mainly due to the regular passage through the South Atlantic Anomaly (SAA) and other Earth orbit perturbations \citep{auv09}. Since this passage occurs twice in a sidereal day, the spectral window (see the inset in Fig.~\ref{spectralwin}) presents a structure composed of a first peak (4.6\% of the amplitude) close to $2.007$~d$^{-1}$ and a second one (1.4\% of the amplitude) close to $4.014$~d$^{-1}$. Moreover, gaps due to the orbital period of the satellite (6184~s) produce other structures with peaks around $13.972$~d$^{-1}$ and their harmonics (Fig.~\ref{spectralwin}). The percentage of lost datapoints because of environmental conditions is only 9.6\% for HD\,47129. In addition, the spectral window exhibits a 89.0\% peak at $f = 2699.764$~d$^{-1}$. This corresponds to the sampling regularity having the highest frequency. Consequently, a pseudo-Nyquist frequency can be located at $f_{\mathrm{Ny}} = 1349.882$~d$^{-1}$ (leading to a step of $32.003$~s). We use the word ``pseudo'' to point out the fact that the aliasing is not pure.

An interpolation of all flagged points of the light curve, to fill the gaps, cleans the spectral windows of all these peaks but this process introduces systematic effects which could generate erroneous values of the frequencies, affecting the scientific results. Accordingly, we decided not to interpolate among the remaining points of the light curve.

All stars observed in the CoRoT field of view present a global slope in their light curve, with probably an instrumental origin (\citealt{auv09} attributed this drift to the ageing of the CCDs). Even though the light curve of Plaskett's Star does not seem to be affected by it, maybe because of its high flux variability, we decided to remove this long-term trend. In order not to bias the possible orbital or long-term variations present in the light curve, the strictly linear trend appears to be the best choice among the different possible trends. However, as we will show in Section~\ref{corot_result}, the low frequencies, close to 0, still appear in the semi-amplitude spectrum after removing the trend.

\begin{figure*}[htbp]
\centering
\includegraphics[width=15.5cm, bb=32 378 585 677,clip]{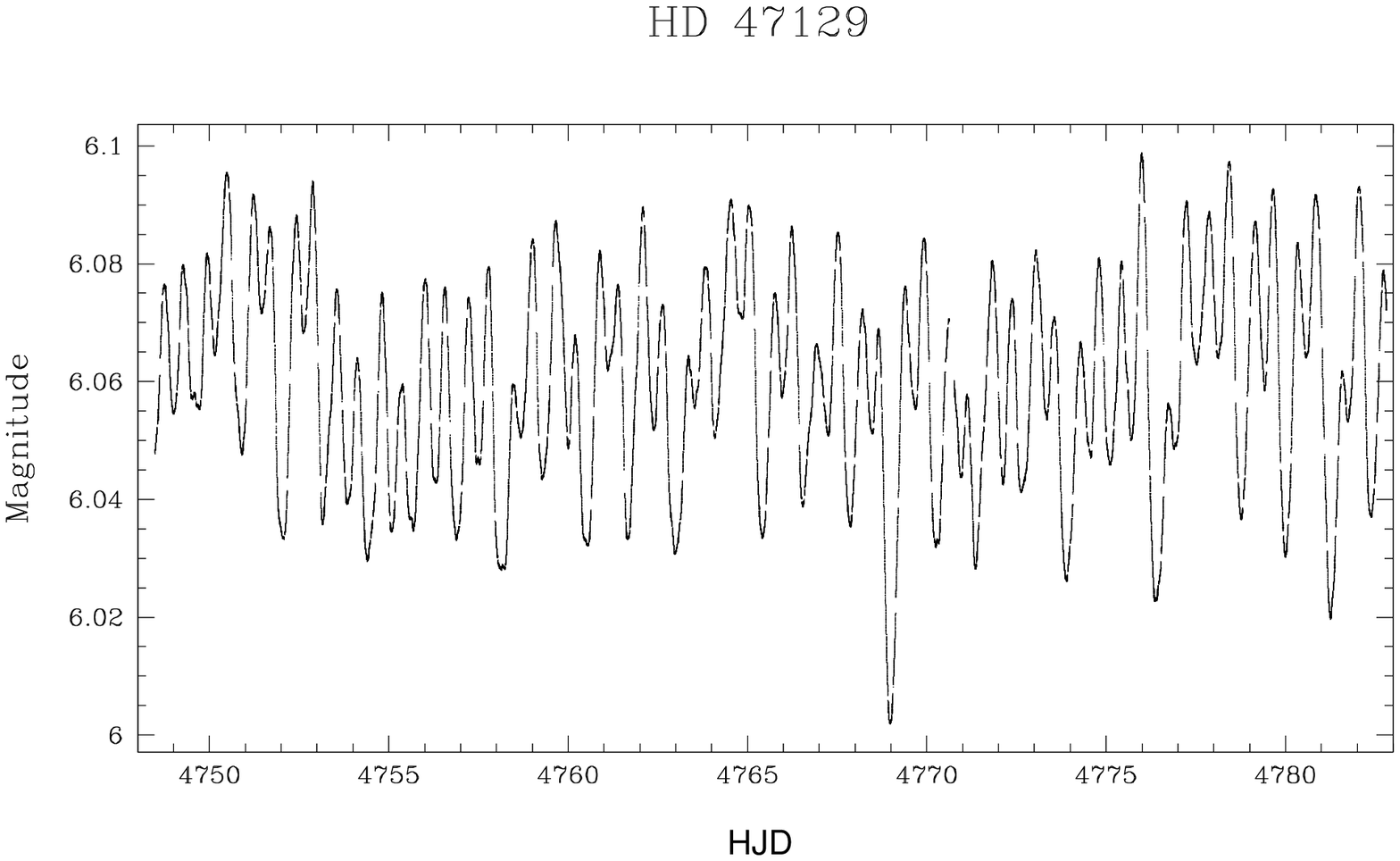}
\caption{Full detrended CoRoT light curve of Plaskett's Star, containing 79896 points, observed over about 34.33~days and converted to magnitude. All gaps present in the data are due to the flagged points described in Sect.~\ref{corot_ana}.\label{corot_curve}}
\end{figure*}

Finally, the CoRoT light curve has also been converted to magnitude from the expression $m = -2.5 \log(F)+C$ where $F$ gives the CoRoT flux expressed in e$^{-}$\,s$^{-1}$ and $C$ represents a calibration constant. We have estimated this constant at a value of $23.09 \pm 0.01$~mag after having compared the CoRoT magnitude to that in the V-band quoted by \citet{lin08}. We note that this conversion in magnitude does not change the decomposition in frequencies reported below. The final version of Plaskett's Star light curve (Fig.~\ref{corot_curve}) is composed of 79896 points.

\section{Analysis of Plaskett's Star CoRoT light curve}\label{corot_result}
\subsection{Looking for periodic structures}\label{freqanalysis}

The CoRoT light curve confirms that Plaskett's Star is not an eclipsing binary although we detect variations that could be linked to the orbital motion of the binary system ($P_{\mathrm{orb}} = 14.39625$~days, see \citealt{lin08}). The light curve also presents large amplitude peak-to-peak variations on shorter time scales. In order to perform a complete investigation of the frequencies present in the CoRoT light curve, we applied a Fourier analysis based on the Heck, Manfroid \& Mersch (hereafter HMM) technique (\citealt{hec85}, revised by \citealt{gos01}). We emphasize that this method is comparable to the Ferraz-Mello one \citep{fm81} and is specially designed to handle time series with unevenly spaced data. Moreover, its mathematical expression for the power spectrum has the advantage of correcting some deficiencies of other methods such as the one of \citet{sca82}. The Fourier technique of HMM, used here, is equivalent, at each individual frequency, to a least-squares fit of a sine function \citep{gos01}. In the present paper, we will actually use the semi-amplitude spectrum instead of the classical power spectrum, implying directly that the ordinates represent the amplitude term in front of the sine function.

\begin{figure*}[htbp]
\centering
\includegraphics[width=16cm, bb=22 400 575 679,clip]{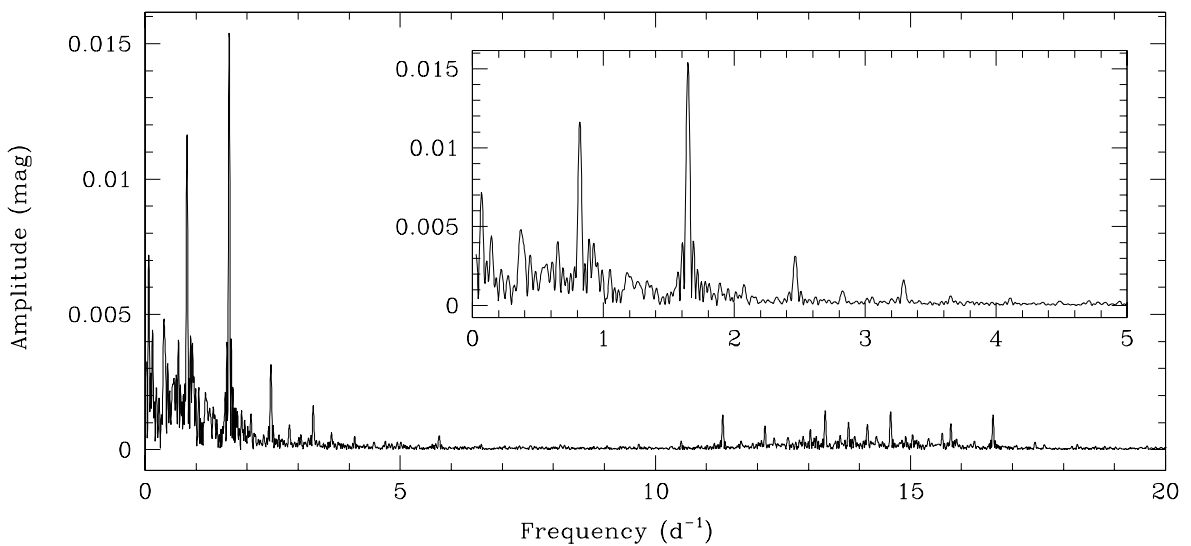}
\caption{Semi-amplitude spectrum of the light curve of Plaskett's Star computed by the HMM method from the unflagged CoRoT data. We clearly see the aliases due to the orbital period of the satellite near 11--17~d$^{-1}$. The inset shows a zoom-in on the semi-amplitude spectrum in the low-frequency domain.\label{sas}}
\end{figure*}

The semi-amplitude spectrum (Fig.~\ref{sas}) clearly shows that most of the power is concentrated in the $f \leq 6.0$~d$^{-1}$ frequency domain and that aliases, generated by the satellite orbital cycle, are present between 11.0 and 17.0~d$^{-1}$. Although we removed the linear trend from the dataset, there remains signal close to frequency $0.0$~d$^{-1}$. The spectrum is dominated by a set of frequencies distributed according to a regular pattern including the fundamental frequency at about $0.82$~d$^{-1}$ and its possible harmonics at about $1.64$~d$^{-1}$, $2.46$~d$^{-1}$, $3.28$~d$^{-1}$ and $4.10$~d$^{-1}$. The $2.007$~d$^{-1}$ aliases of the two first (main) frequencies are also visible at $f = 2.83$~d$^{-1}$ and $3.65$~d$^{-1}$. Besides these frequencies, the semi-amplitude spectrum confirms the presence of $f = 0.07$~d$^{-1}$, $f=0.14$~d$^{-1}$ and probably $f = 0.21$~d$^{-1}$, i.e., the orbital frequency and its harmonics which constitute a second set of frequencies. Further outstanding peaks are also present at $f \sim 0.35$~d$^{-1}$, $f \sim 0.65$~d$^{-1}$ and $f \sim 0.95$~d$^{-1}$. The first one having a wider peak that expected from the time basis, it could actually be a blend of several frequencies (at least two, or one plus an alias of $f = 1.64$~d$^{-1}$). 

\begin{figure}[htbp]
\centering
\includegraphics[width=9cm, bb=18 167 570 695,clip]{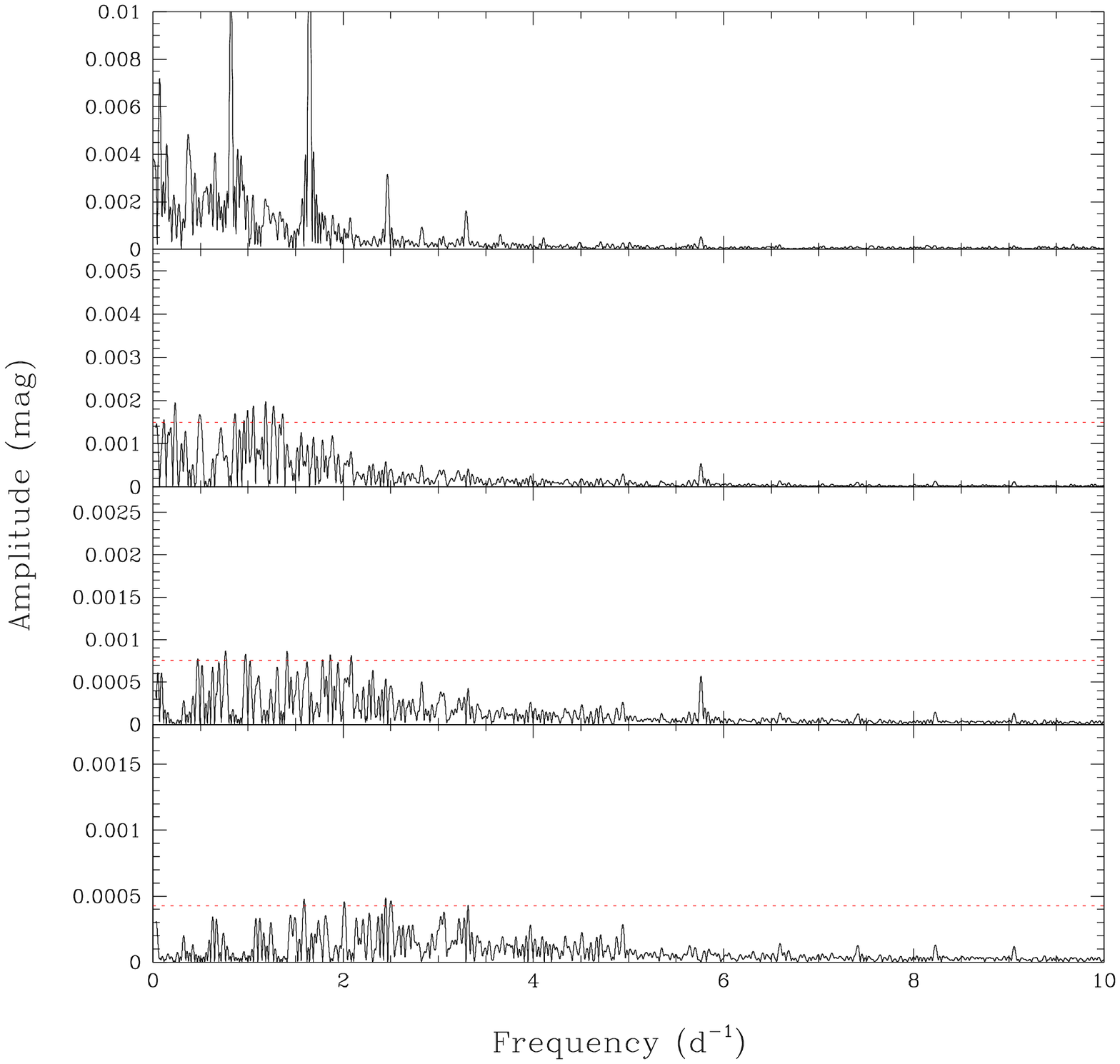}
\caption{For comparison, the top panel shows the semi-amplitude spectrum before prewhitening, the subsequent panels represent the semi-amplitude spectrum after prewhitening 20, 40 and 60 frequencies, respectively. Red lines exhibit the critical level at the significance level of 0.01 under a null hypothesis of white noise.\label{corot_reconst}}
\end{figure}

In order to analyse the time series in a more systematic way, the in-depth determination of the different frequencies was done in two steps. First, we proceeded to a classical iterative prewhitening of the signal, frequency by frequency. At each step, the semi-amplitude spectrum is computed and the frequency value is selected by the position of the highest peak in the spectrum. The amplitude and the phase, corresponding to this frequency, are directly computed from the Fourier function. We however initiated the process by removing the set of 5 frequencies composed, notably, of the two highest peaks found in the periodogram ($f = 0.82$ and $f = 1.64$~d$^{-1}$) and the frequencies belonging to this set. Since the frequency $f = 0.82$~d$^{-1}$ presents possible harmonics, we decided to perform the fit on all of them together, taking into account the existing correlation between the power at the different frequencies. We thus designed a generalized periodogram giving at each frequency the power included in a frequency and {\em its harmonics}. The method is based on a particular case of the extension of the HMM method proposed by \citet{gos01}. The generalized periodogram exhibits the $f = 0.82$~d$^{-1}$ peak characterized by the natural width, thus further confirming the fact that this sequence of peaks are indeed true harmonics. These frequencies were removed altogether from the data. Next, the semi-amplitude spectrum has been computed again in order to detect the new highest peak to remove. The analysed data are prewhitened for this frequency and we repeated this process until reaching the noise level of the data. Fig.~\ref{corot_reconst} compares the semi-amplitude spectra of the data and of the data prewhitened for the first 20 frequencies, the first 40 and the first 60 ones. 

Although efficient, the iterative removing of several frequencies, one by one, by using the basic HMM method, can not be the final procedure. Since the light curve presents gaps and the sampling is irregular, the height of a peak is dependent on the height of other peaks in the periodogram. Therefore, we need to fit all the listed frequencies together. For this purpose, we used the extension of the HMM Fourier method to the multifrequency adjustment, a high order Fourier method introduced by \citet[][their Eq.~A13 to A19]{gos01}. Because of an excessive computation time, we applied the multifrequency analysis to binned data to be able to deal with some amount of frequencies at the same time. Indeed, since no peak is clearly dominant above $10$~d$^{-1}$, we have binned the data with a step of one twentieth of a day, i.e., with a pseudo-Nyquist frequency of $f_{\mathrm{Ny}} = 10$~d$^{-1}$, reducing the CoRoT light curve into a set of 684 points. The multifrequency algorithm thus takes into account the mutual influence of the different frequencies and investigates, inside the natural widths of the individual selected peaks, several positions, i.e., refined values for each of the frequencies in order to find the current best fit.

The results obtained by both methods confirm the detection, in the semi-amplitude spectrum, of a first structure composed of a main frequency and six harmonics. The fundamental peak is situated around 0.823~d$^{-1}$ and its harmonics around $1.646$~d$^{-1}$, $2.469$~d$^{-1}$, $3.292$~d$^{-1}$, $4.115$~d$^{-1}$, $4.938$~d$^{-1}$ and $5.761$~d$^{-1}$, respectively. Furthermore, we also clearly detect the peak corresponding to the orbital period of Plaskett's Star ($f \sim\!0.069$~d$^{-1}$) as well as two harmonics of this frequency. We also emphasize the detection of two frequencies, actually unresolved, at $f = 0.368$~d$^{-1}$ and $f = 0.399$~d$^{-1}$. 

The set of frequencies, computed in that way, represents the final set (listed in Appendix). Actually, we stop the iterative multifrequency procedure when the noise level is reached (statistically). This critical level is particularly difficult to estimate and constitutes the main topic of the next section.

\subsection{The noise properties and significant frequencies}\label{noisesec}

\begin{figure}[htbp]
\centering
\includegraphics[width=9cm, bb=38 408 568 674,clip]{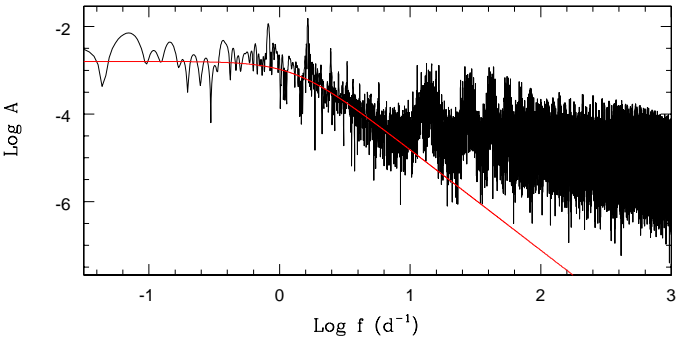}
\caption{Log-log plot of the semi-amplitude spectrum of Plaskett's Star. The red curve represents the fitted function (Eq.~\ref{eq1}).\label{loglogplot}}
\end{figure}

From the analysis of the time series, illustrated in Figs.~\ref{sas} and~\ref{corot_reconst}, we conclude that there is a clear excess of power at low frequencies. It represents either the reality of the underlying deterministic signal or the fact that the signal is partly made of red noise, i.e., is partly stochastic. Visible on the log-log plot (Fig.~\ref{loglogplot}) of the periodogram of Plaskett's Star, this excess of power at low frequencies ($\log f \leq 0.6$) can be, for example, described, as suggested by \citet{sta02}, by using a function of the form: 
\begin{equation}\label{eq1}
P(f) = \frac{C}{1+(2 \pi \tau f)^{\gamma}}
\end{equation}
with $\gamma$ being related to the slope of the linear part and $\tau$ being an estimation of the mean duration of the dominant transient structures in the light curve. A least square fit of the semi-amplitude spectrum in the low-frequency domain yields parameters $\gamma=2.3$ and $\tau=0.12$~d$^{-1}$. Assuming an origin at least partly intrinsic to the star, several processes could explain the behaviour of the stochastic component of the signal. It could be either due to an onset of clumping at the stellar photosphere or, if we make the parallelism with the works of \citet{har85} and \citet{aig04} in helioseismology, to some kind of granulation. Indeed, \citet{can09} suggested that the convection zone induced by the iron opacity bump can have an impact on the stellar surface behaviour and thus could be responsible for the existence of red noise. \citet{bel10} further suggested that this iron convection zone could generate stochastically excited modes in massive stars. Finally, other alternatives are also mentioned to explain this origin as the non-linearity \citep[for more details, see e.g.,][]{per09}. 

Despite the possible presence of red noise, it is thus important to check whether peaks detected in the semi-amplitude spectrum might result from random variations rather than representing a periodic signal. However, the evaluation of the absolute significance of a peak in a periodogram of a time series with an uneven sampling is a controversial problem \citep[e.g.,][]{rau08,fre08}. Furthermore, the detection threshold level will be affected by this assumption of coloured noise, making its estimate more difficult. As a first order indication, we have thus applied a statistical test to establish whether the frequencies are significant, especially for those with smaller amplitudes. The probability that the power at several inspected frequencies exceeds a threshold $z$ under the null hypothesis of a stochastic process of variance $\sigma^2_{f}$ (function of frequencies) is given by the empirical formula of \citet{gos07}:  
\begin{equation}
\displaystyle{\mathrm{Prob}[Z_{\mathrm{max}}>z]= 1-(\mathrm{e}^{-\mathrm{e}^{-0.93 z + \ln {(0.8 N_0)}}})}
\end{equation}
where $Z_{\mathrm{max}} = \displaystyle{\max_{0< f < f_{\mathrm{Ny}}} Z(f)}$, $z = P_f/\sigma_f^2$ with $P_f$ the power in the power spectrum at the frequency $f$ which is related to the semi-amplitude by $P_f = N_0 A^2_f/4$ and $N_0$ represents the number of datapoints.
Although this expression is only adapted to even sampling, it remains a good approximation (since no better alternative exists) for uneven ones, especially for the CoRoT sampling which is more reminiscent of a regular but gapped one. We adopted the number of datapoints in the binned light curve, i.e., $N_0= 684$. A first estimation of the significance level can be made by adopting a null hypothesis of white noise (i.e., $\sigma^{2}_f$ not depending on $f$). In this case, we detect 79 frequencies (60 frequencies) at a significance level of 0.01 (0.001) or lower. A second approach is to adopt a more complex stochastic process as null hypothesis. In this case, we defined a stochastic distribution law by assuming an empirical function based on white noise above $f = 6.3$~d$^{-1}$ and on red noise defined by the continuum in the power spectrum below $f = 6.3$~d$^{-1}$. We have normalized the above fitted function in order to define a properly scaled $\sigma^{2}_f$ function. Under this refined null hypothesis of red noise, the number of significant frequencies decreases to about 43 (38) at a significance level of 0.01 (0.001). Table~\ref{table1} exhibits the details of the significant frequencies detected under the null hypothesis of red noise. The columns represent the sequence number of the frequency, the frequency itself, its amplitude, the significance level under the null-hypothesis of white and red noise and some comments, respectively.

\begin{figure}[htbp]
\centering
\includegraphics[width=9cm, bb=40 176 555 675,clip]{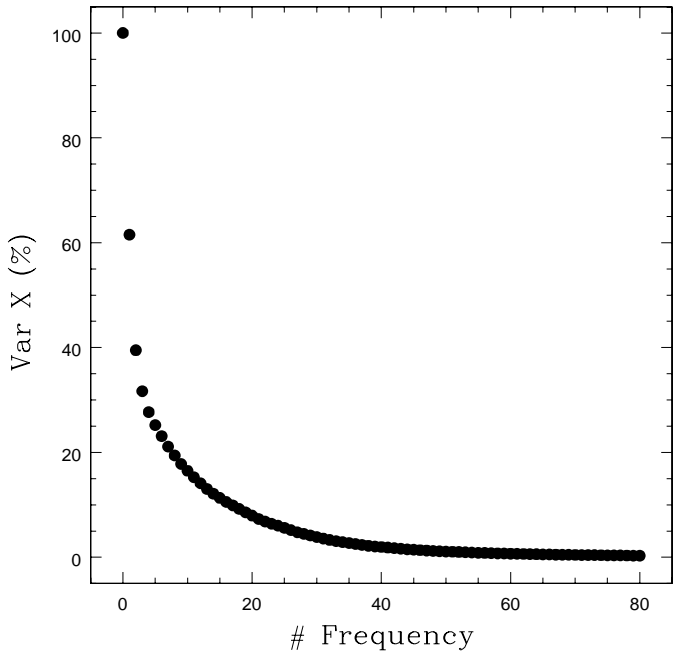}
\caption{Evolution of the relative variance of the prewhitened data as a function of the number of frequencies already detected by the multiperiodic algorithm.\label{variance}}
\end{figure}

If we look at the evolution of the data variance with the iterative prewhitening process (Fig.~\ref{variance}), we note that there is no jump in the curve, which underlines the difficulty to define a threshold. Indeed, both structures formed by the fundamental peaks at $f = 0.823$ and $f=0.069$~d$^{-1}$ already explain 72\% of the variance of the CoRoT light curve and, after 20 frequencies, the total variance of the prewhitened observations strongly decreased. 

The decomposition of the signal into frequencies is a formal process that does not necessarily represent the physical truth. Moreover, it neglects non linearities and complex phenomena. It is thus impossible to derive a list of frequencies free of contaminations. However, the truly existing frequencies (orbit, possible pulsations,...) should be part of this derived list ({\bf reported} in the Appendix). We also decided to use an additional method to contribute to the determination of the reliable frequencies although this method is not perfect either. We {\bf split} the data into two halves: a first one gathering the observations taken between $HJD = 4748.48$ and $4765.63$ and a second one taking into account the data collected between $HJD = 4765.63$ and $4782.82$. We applied the frequency research on each sample separately by using the same technique as above and we directly compared both frequency lists. This method allowed us to put forward a list of 30 frequencies common to both dataset.  

The error estimate for the frequencies (listed in Appendix) is obtained from the expression given by \citet{luc71} and {\bf \citet{mon99}}: 
\begin{equation}
\epsilon(f) = \frac{\sqrt{6}}{\pi} \cdot \frac{\sigma}{\sqrt{N_0} \Delta T} \cdot \frac{1}{A}
\end{equation}
where $\sigma$ is the standard deviation of the residuals, $A$ expresses the semi-amplitude associated to the frequency, $\Delta T$ represents the observation span of time and $N_0$, the number of points of the binned sample (684). However, these formal errors are often underestimated compared to the actual ones \citep{deg09b}. The list of frequencies given in the Appendix contains 150 values and were stopped at the last significant ($<$0.01) frequency. It is noted that the first 19 frequencies are highly significant against both null hypothesis (white and red noise) but also present in both partitional datasets.

\begin{table}
\caption{Table of main frequencies}\label{table1}
\begin{tabular}{lccccc}
\hline\hline
$n$ & Freq. & Semi-Ampl. & SL$_{\mathrm{White}}$ & SL$_{\mathrm{red}}$ & Comments\\
& ($d^{-1}$) & (mmag) & & &\\
\hline
2 & 1.646 & 15.413 & 0.000 & 0.000 & $2f_2$\\
1 & 0.823 & 11.373 & 0.000 & 0.000 &\\
8 & 0.069 &  6.660 & 0.000 & 0.000 &\\
11 & 0.368 &  5.026 & 0.000 & 0.000 &\\
12 & 0.650 &  3.732 & 0.000 & 0.000 &\\
9 & 0.139 &  3.410 & 0.000 & 0.000 & $2f_3$\\
3 & 2.469 &  3.124 & 0.002 & 0.000 & $3f_2$\\
15 & 0.932 &  3.121 & 0.000 & 0.000 &\\
16 & 0.441 &  3.099 & 0.000 & 0.000 &\\
14 & 0.888 & 3.069 & 0.000 & 0.000 &\\
13 & 0.399 & 3.007 & 0.000 & 0.000 &\\
17 & 0.542 & 2.620 & 0.000 & 0.000 &\\
18 & 0.799 & 2.567 & 0.000 & 0.000 &\\
21 & 0.572 & 2.524 & 0.000 & 0.000 &\\
20 & 0.607 & 2.406 & 0.000 & 0.000 &\\
22 & 1.212 & 2.042 & 0.000 & 0.000 & \\
23 & 1.185 & 1.904 & 0.000 & 0.000 &\\
28 & 1.264 & 1.877 & 0.000 & 0.000 & \\
24 & 1.056 & 1.864 & 0.000 & 0.000 &\\
30 & 1.334 & 1.786 & 0.000 & 0.000 & \\
27 & 1.000 & 1.616 & 0.000 & 0.007 &\\
32 & 1.359 & 1.598 & 0.000 & 0.000 &\\
4 & 3.292 & 1.533 & 1.000 & 0.000 & $4f_2$\\
33 & 1.558 & 1.362 & 0.000 & 0.000 & \\
37 & 1.890 & 1.124 & 0.000 & 0.000 & \\
38 & 1.686 & 1.079 & 0.000 & 0.000 &\\
46 & 1.865 & 0.860 & 0.000 & 0.000 &\\
47 & 2.085 & 0.834 & 0.000 & 0.000 & \\
63 & 2.312 & 0.576 & 0.001 & 0.001 &\\
7 & 5.761 & 0.525 & 1.000 & 0.000 & $7f_2$\\
68 & 2.825 & 0.510 & 0.001 & 0.000 & Alias + $f_2$?\\
69 & 2.446 & 0.496 & 0.001 & 0.004 & \\
71 & 2.503 & 0.463 & 0.002 & 0.009 & \\
74 & 3.314 & 0.425 & 0.004 & 0.000 & \\
76 & 3.061 & 0.386 & 0.013 & 0.000 & \\
5 & 4.115 & 0.350 & 1.000 & 0.000 & $5f_2$\\
6 & 4.938 & 0.275 & 1.000 & 0.000 & $6f_2$\\
95 & 3.969 & 0.256 & 0.110 & 0.000 & \\
102& 4.508 & 0.225 & 0.197 & 0.000 & \\
106& 4.709 & 0.204 & 0.366 & 0.000 & \\
115& 4.881 & 0.182 & 0.411 & 0.000 & \\
148& 5.646 & 0.133 & 0.298 & 0.000 & \\
150& 5.698 & 0.127 & 0.412 & 0.001 & \\

\hline
\end{tabular}
\end{table}

\section{Discussion}\label{corot_disc}
\subsection{The inclination of the orbit of Plaskett's Star}\label{corot_disc_1}

Even though the variations between the maximum and minimum observed on the entire run are close to $\sim\!0.1$~mag, the unprecedented quality of the CoRoT light curve gives us enough information to attempt a study of the inclination of Plaskett's Star. For this purpose, we use the \texttt{NIGHTFALL} program\footnote{For details, see the \texttt{NIGHTFALL} User Manual by \citet{wic98} available at the URL:~\texttt{http://www.hs.uni-hambourg.de/DE/Ins/\\Per/Wichmann/Nightfall.html}} to fit the variations due to the orbital motion. This program is based on a generalised Wilson-Devinney method assuming a standard Roche geometry.

We have used the information from our Fourier analysis by combining the amplitudes and phases of the peaks detected at the orbital frequency and its first two harmonics ($f = 0.139$ and $f = 0.208$~d$^{-1}$). The photometric variations of HD\,47129 tied to the orbital cycle have a peak-to-peak amplitude of about 19\,mmag. The orbital light curve displays a broad minimum roughly centered on phase 0.0 (the conjunction with the primary star being in front)\footnote{With respect to the time of phase 0.0 defined in \citet{lin08}, the minimum of the light curve is shifted by about 0.03 in phase. This is somewhat larger than what can be accounted for by the uncertainty on the orbital period quoted by Linder et al. Here we have chosen to take phase 0.0 at the time of the minimum of the light curve.}. This is followed by a broad maximum around phase 0.3, whilst there is no clear secondary minimum, but rather a kind of plateau before the brightness decreases towards phase 0.0. These variations are clearly not due to grazing eclipses and cannot be explained by the sole effect of ellipsoidal variations either. Indeed, pure ellipsoidal variations in a system with a circular orbit, such as Plaskett's Star, would produce a light curve with two equally deep minima centered on phases 0.0 and 0.5. We note that ellipsoidal variations in an eccentric binary could potentially account for the observed light curve (see e.g.\ the case of Tr\,16-112, \citealt{rau09}). However, the radial velocity curves of Plaskett's Star (\citealt[][]{lin08} and references therein) provide no evidence whatsoever for a non-zero eccentricity. An alternative possibility to account for the shape of the orbital light curve of Plaskett's Star is to assume a configuration where one of the stars has a hot spot on its surface. Indeed, the presence of such a hot, and hence bright, region could counterbalance the small ellipsoidal variations. 
\begin{figure}[htbp]
\begin{center}
\resizebox{9cm}{!}{\includegraphics{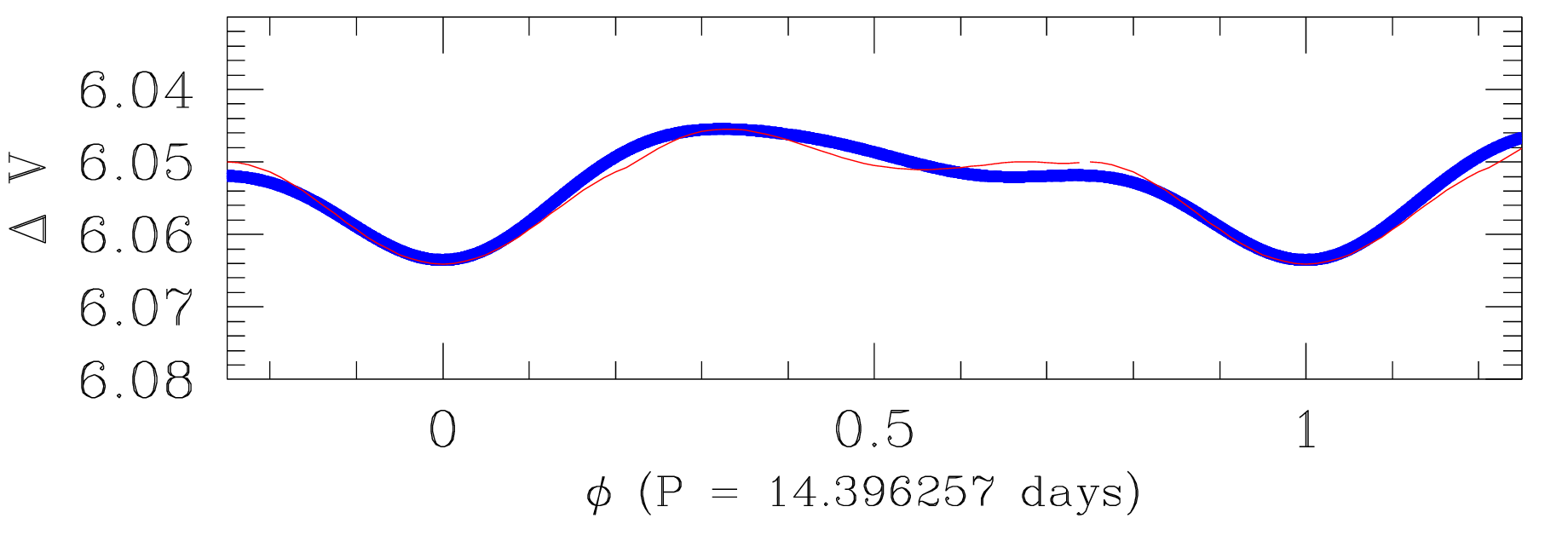}}
\end{center}
\caption{Example of a model fit of the orbital part of the light curve of Plaskett's Star. The parameters of the model (red thin curve) are $i = 67^{\circ}$, fill$_P$ = 0.54, fill$_S$ = 0.450, longitude of the spot = $350^{\circ}$, radius of the spot = $7^{\circ}$ and temperature ratio (spot/primary star) = 1.97.\label{fig-1}}
\end{figure}
We fit the light curve of Plaskett's Star with the following assumptions: the mass ratio as well as the temperatures of the stars are fixed according to the results of \citet{lin08}. A priori, the hot spot could be located on the surface of either star. However, to fit the observed light curve, a spot on the secondary, would have to be on the rear side of the star (i.e.\ turned away from the primary), whilst a spot on the primary star would be located on the side of the star facing the secondary. The latter configuration seems more plausible, especially in comparison with the wind interaction model of \citet{lin08}. We therefore assume that the hot spot lies on the surface of the primary. The secondary is taken to rotate asynchronously with a rotation period one quarter of the orbital period, although we stress that this assumption has little impact on our results.  
At first, the inclination, the Roche lobe filling factors of both stars, the longitude of the hot spot, its radius and temperature ratio are taken as free parameters. The small amplitude of the light curve prevents us, however, from deriving strong constraints on the values of all these parameters. This dilemma can be illustrated by considering for instance the orbital inclination. As a matter of fact, when systematically exploring parameter space, we find equally good fits for values of $i$ ranging from 30 to 80$^{\circ}$. Obviously to reproduce the observed light curve with $i = 80^{\circ}$ requires extremely low filling factors of at least one of the stars and this is unlikely since the spectroscopic investigation revealed a luminosity ratio of about 1.9 for stars with essentially identical temperatures (see \citealt{lin08}). Conversely, the lower inclination solutions yield unrealistically large masses. When we constrain the filling factors in such a way as to reproduce the spectroscopic luminosity ratio, we find the best fit for an inclination of about $67^{\circ}$. Whilst this number seems quite `reasonable', we stress that it must be taken with caution. We also attempt to fit the ellipsoidal variations individually (i.e., limiting ourselves to the double of the orbital frequency and not accounting for a spot) but, once again, no exact determination of the inclination has been possible. Indeed, the fit of this parameter is directly dependent on the filling factor of both stars. 

For all the solutions that we find, the longitude of the bright spot is found between $345$ and $355^{\circ}$ ($0^{\circ}$ corresponding to the direction of the secondary star and $90^{\circ}$ indicating the direction of the motion of the primary). Furthermore, this spot has a temperature ratio of about 1.9 with respect to the surface temperature of the primary. The radius of the spot is found to be about 8 -- $10^{\circ}$ for most solutions, except for those with very high orbital inclinations and low filling factors.

A major problem concerning Plaskett's Star remains the distance at which the binary system is situated. Indeed, whatever the configuration used to fit the light curve, the parameters used by \texttt{NIGHTFALL} indicate that the star should be located between 2.0 and 2.2~kpc. The increase of the distance would imply that Plaskett's Star does not belong to the Mon~OB2 association, as was already suspected by \citet{lin08}. These authors indeed found a discrepancy between the spectral types of both stars and their dynamical masses. If we compare the systemic velocity of Plaskett's Star, estimated at $30.6 \pm 1.8~\kms$ by \citet{lin08}, with the radial velocities of other O-type stars situated in Mon~OB2 association \citep{mah09}, it appears that these values agree between each other and that Plaskett's Star would belong to this association. However, this assumption will only be checked with measurements from the future Gaia mission.

In summary, whilst the light curve does not allow us to establish the inclination of the orbit, we find that its shape is consistent with moderate ellipsoidal variations altered by the presence of a rather bright spot on the primary star, facing the secondary. This spot is very probably related to the wind interaction between the two stars and could actually be due to shock-heated material that is cooling whilst it flows away from the stagnation point.

\subsection{Structures present in the light curve}\label{corot_disc_2}

The multiperiodic analysis of the CoRoT light curve allowed us to detect a significant structure composed of a fundamental frequency ($f = 0.823$~d$^{-1}$) and its six harmonics but the nature of this signal is not yet established.

In a binary system such as Plaskett's Star, a possible cause of spectroscopic or photometric variability could be the asynchronous rotation. Indeed, a binary system is in synchronous rotation when the angular rotation velocity $\omega$ and the angular velocity of orbital motion $\Omega$ are equal. In HD\,47129, this is clearly not the case: adopting the projected rotational velocities and stellar radii from \citet{lin08}, we estimate rotational periods of about 10.3 ($0.1$~d$^{-1}$) and 1.7\,days ($0.6$~d$^{-1}$) for the primary and secondary, respectively. As a result of asynchronous rotation, tidal interactions may create non-radial oscillations \citep{wil02,mor05,koe10}. These oscillations produce a pattern of azimuthal velocity perturbations superposed on the unperturbed stellar rotation field. These tidal flows lead to the dissipation of energy due to the viscous shear, thereby impacting on the surface temperature (and hence brightness) distribution.
The azimuthal velocity components of tidal interactions are expected to produce strong line profile variability that resembles the typical signature of non-radial pulsations (bumps in the line profile that migrate from the blue wing of the spectral lines to the red wing). The most spectacular effects are expected in eccentric binary systems and this model was successfully used to explain the line profile variability of the eccentric B-type binary $\alpha$\,Virginis \citep{mor05,koe10}. Simulations of the influence of the tidal effects on the radial velocities were shown by \citet{wil02}.

In a binary system with a circular orbit, the non-synchronous rotation induces variations of the radius of the non-synchronous rotating component with a super-orbital period that is longer than both the orbital and the rotational period \citep{mor05}. Whilst these super-periods could leave a photometric signature, they should not be visible in the line profile variability which is mostly dominated by the azimuthal velocity component. \citet{mor05} found that, in the circular non-synchronous case, the line profiles should only display a few, rather broad bumps. The skewness of the line profile and its radial velocity (compared to the unperturbed profile) should display two maxima and minima per orbital cycle. The corresponding frequency should thus be twice the orbital frequency which is different from $f = 0.823$~d$^{-1}$. 

We then consider that the structure composed of 7 frequencies could be due to the rotation of one component of the binary system. In this context, we clearly see by adopting the rotational periods from \citet{lin08} that this set of frequencies is unlikely to correspond to the rotation of one star. A similar conclusion is expected if we take into account the projected rotational velocity evaluated by \citet{lin08} for both stars at about $75$ and $310~\kms$ for the primary and secondary component, respectively. Indeed, \citet{Bag92} derived the radii from the luminosities and the effective temperatures of both stars. They found values close to $14.7~\rsun$ and $9.7~\rsun$ for the primary and secondary stars, respectively. We estimate a rotational period of about $9.9$ and $1.6$~days for the primary and secondary stars, respectively, corresponding to rotational frequencies of about $0.1$ and $0.6$~d$^{-1}$. Alternatively, if we adopt the theoretical values of the radii quoted by \citet{mar05} as a function of the spectral type of each component, we obtain $21.1~\rsun$ and $14.2~\rsun$ for the primary and secondary stars, respectively, by assuming that the system is composed of an O8\,I and an O7.5~III star. These values yield a rotational period of $14.2$~days for the primary and $2.4$~days for the secondary. These results suggest that the primary could be in synchronous rotation while this is clearly not the case for the secondary, which would have a rotational frequency of about $0.42$~d$^{-1}$. In both cases, it thus appears that the fundamental frequency found in the Fourier analysis is not representative of a rotational motion, except if we could invoke a spotted surface for the secondary. But, in this case, the lack of odd multiple of the rotation frequency is surprising. By assuming a rotational frequency of $f = 0.823$~d$^{-1}$, we would obtain a radius of about $7.2~\rsun$ for the secondary star, which seems unrealistic for such stars. However, it is possible that the frequency at $f = 0.368$~d$^{-1}$ or $f = 0.399$~d$^{-1}$ is representative of the rotational period of the secondary component. In the latter case, the equatorial radius of the secondary star could be estimated close to $16.7$ or $15.4~\rsun$, respectively.

Next, we suppose that this group of frequencies could arise from wind interactions. \citet{wig92} reported on variations of the strength of the high-velocity wings of the H$\alpha$ and He\,{\sc i} $\lambda$\,6678 emission line profiles. The total equivalent width of the blue and red wings of H$\alpha$ was found to display a roughly recurrent modulation with a time scale of 2.82\,days ($0.35$~d$^{-1}$), although the sampling of the spectra of \citet{wig92} is admittedly not sufficient to characterize such rapid variations. These authors attributed the variations to instabilities of a radiatively cooling wind-wind interaction. It is interesting to note that this frequency is very close to the ``broad peak'' detected in the analysis of our photometric data (Sect.~\ref{freqanalysis}). It has to be stressed though that the {\it XMM-Newton} observations of \citet{lin06} did not reveal any indication of X-ray variability that could be related to the existence of such instabilities in the wind interaction zone.

Finally, we consider that these frequencies could be generated by non-radial pulsations. Indeed, massive stars are composed of a convective core and a radiative envelope, and gravity as well as acoustic modes can be excited \citep{aer10}. Generally, for massive stars situated on the main-sequence band, the p-modes propagate with periods of a few hours while the g-modes have longer periods, of order of days. However, in the case of Plaskett's Star, \citet{lin08} concluded at an evolved system in a post Roche lobe overflow evolutionary stage. The inner structure of such stars is different from main-sequence stars, which decreases the g-modes periods and, at the same time, increases the p-modes periods \citep{han05}. Another family of modes, the strange modes \citep{sai98}, could also play a role in the interpretation of this structure of frequencies. These modes are known to have a propagation zone near the surface of massive stars.
The exact determination of these modes requires theoretical models to predict variations produced by non-radial pulsations in a rapidly rotating massive star as well as an intense spectroscopic monitoring to characterize the variability of the line profiles. However, for Plaskett's Star, massive main-sequence star models computed with ATON \citep{ven08} and an initial metallicity of $0.015$, a range of mass between $30$ and $70~\msun$ and different mass-loss rates show that non-radial pulsations with frequency of the order of $0.8$~d$^{-1}$ can indeed be found. This frequency could be generated by modes with $l = 2, 3$ or $4$.

In summary, we conclude that the alternative scenarios linked to rotation or to winds, in both stars, fail to explain the structure of the Fourier spectrum, especially the dominant frequency and its set of harmonics. The assumption of multiperiodic non-radial pulsations remains the most plausible one to understand the low-frequency variability detected in the CoRoT light curve of Plaskett's Star.

\section{Conclusions}\label{corot_concl}
We presented the photometric analysis of the very massive binary system, HD\,47129, observed by CoRoT during the second run SRa02 ($\sim \! 34.33$~days). The light curve shows indications of intrinsic variations of the stars, providing evidence of a large number of frequencies. We extracted a total of about 43 frequencies (Table~\ref{table1}), significant under a null hypothesis of red noise, by using two different techniques based on a standard prewhitening and a multiperiodic algorithm. We emphasize that all the frequencies, reported in the present paper, have not necessarily a physical meaning.

This analysis highlighted a group consisting of one fundamental frequency ($0.823$~d$^{-1}$) and its six harmonics. In addition, a second structure formed by the orbital frequency ($0.069$~d$^{-1}$) and its two harmonics allowed us to investigate the light variation due to orbital effects. This analysis revealed the presence of a hot spot, probably located near the primary star, facing the secondary component and having an origin in the colliding wind interaction zone. However, the determination of the exact value of the system inclination turned out to be ambiguous. A third structure, formed by a much wider peak and thus certainly composed of two frequencies at $0.368$~d$^{-1}$ and $0.399$~d$^{-1}$, has also been detected. One of these frequencies could be related to the rotation of the secondary even though \citet{wig92} rather suggested an origin in the winds of the stars.

The future work will be devoted to the interpretation of these frequencies in terms of asteroseismology in order to remove an additional part of mystery concerning Plaskett's Star. Moreover, a study of the line profiles either of the secondary or of the primary component, obtained from an intense high-resolution spectroscopic campaign, could provide further constraints on the properties of the pulsation modes. Indeed, the broad and shallow line profiles of the secondary star favor this detection although the orbital motion and the presence of the primary star will certainly render this task more difficult.

\begin{acknowledgements}
We thank the CoRoT team for the acquisition and the reduction of the CoRoT data. This present work was supported by the F.N.R.S (Belgium), the PRODEX XMM/Integral contract (Belspo), Gaia DPAC Prodex and the Communaut\'e fran\c caise de Belgique -- Action de recherche concert\'ee -- A.R.C. -- Acad\'emie Wallonie-Europe. The research leading to these results has also received funding from the European Research Council under the European Community's Seventh Framework Programme (FP7/2007--2013)/ERC grant agreement n$^\circ$227224 (PROSPERITY), from the Research Council of K.U.Leuven (GOA/2008/04), and from the Belgian federal science policy office (C90309: CoRoT Data Exploitation). M.G. and A.N. thank J. Montalban and P. Ventura for help in computation aspects of massive stars.

\end{acknowledgements}

\bibliographystyle{aa}
\bibliography{laurent.bib}

\Online
\begin{appendix}

\longtab{2}{
\begin{landscape}
\begin{longtable}{rcccccccccc}
\caption{\label{kstars} Total list of frequencies. The errors given on the frequencies, semi-amplitudes and phase correspond to the 1--$\sigma$ error. The $T_0$ of the phase is equal to 2,450,000. The last column marks the frequencies present both in the first part of the light curve and in the second one.}\\
\hline\hline
$n$& Frequency  & Frequency errors & Semi-Ampl. & Semi-Ampl. errors & Phase & Phase errors & Variance & SL$_{\mathrm{white}}$& SL$_{\mathrm{red}}$ & Present in both halves \\
 & (d$^{-1}$) & (d$^{-1}$) & (mmag) & (mmag) & (rad) & (rad) & $\times$ 10$^{-4}$ & & & \\
\hline
\endfirsthead
\caption{continued.}\\
\hline\hline
$n$& Frequency  & Frequency errors & Semi-Ampl. & Semi-Ampl. errors & Phase & Phase errors & Variance & SL$_{\mathrm{white}}$& SL$_{\mathrm{red}}$ & Present in both halves \\
 & (d$^{-1}$) & (d$^{-1}$) & (mmag) & (mmag) & (rad) & (rad) & $\times$ 10$^{-4}$ & & & \\
\hline
\endhead
\hline
\endfoot
1 & 0.823 & 0.001 & 11.373& 0.949 & -1.220&  0.083 & 3.080995 & 0.0000 & 0.0000 &* \\ 
2 & 1.646 & 0.001 & 15.413& 0.842 & -0.622&  0.055 & 2.423326 & 0.0000 & 0.0000 &* \\
3 & 2.469 & 0.003 & 3.124 & 0.601 & -0.747&  0.192 & 1.235682 & 0.0019 & 0.0000 &* \\
4 & 3.292 & 0.006 & 1.533 & 0.589 & 0.220 & 0.384 & 1.185302 & 1.0000 & 0.0000 &* \\
5 & 4.115 & 0.027 & 0.350 & 0.586 & -0.037&  1.673 & 1.174480 & 1.0000 & 0.0000 &* \\
6 & 4.938 & 0.034 & 0.275 & 0.586 & -0.965&  2.130 & 1.173976 & 1.0000 & 0.0000 &* \\
7 & 5.761 & 0.018 & 0.525 & 0.586 & 0.982 & 1.116 & 1.173613 & 1.0000 & 0.0000 &* \\
8 & 0.069 & 0.001 & 6.660 & 0.585 & -1.084&  0.088 & 1.172099 & 0.0000 & 0.0000 &* \\
9 & 0.139 & 0.002 & 3.410 & 0.522 & 1.335 & 0.153 & 0.931321 & 0.0000 & 0.0000 &* \\
10 & 0.208 & 0.008 & 1.069 & 0.505 & -1.284&  0.472 & 0.871496 & 1.0000 & 1.0000 &* \\
11 & 0.368 & 0.002 & 5.026 & 0.503 & -1.213 & 0.100 & 0.865742 & 0.0000 & 0.0000 &* \\
12 & 0.650 & 0.002 & 3.732 & 0.468 & 1.264 & 0.125 & 0.748715 & 0.0000 & 0.0000 &* \\
13 & 0.399 & 0.002 & 3.007 & 0.442 & 0.230 & 0.147 & 0.667486 & 0.0000 & 0.0000 &* \\
14 & 0.888 & 0.002 & 3.069 & 0.421 & -0.513&  0.137 & 0.607357 & 0.0000 & 0.0000 &* \\
15 & 0.932 & 0.002 & 3.121 & 0.403 & -0.309&  0.129 & 0.555454 & 0.0000 & 0.0000 &* \\
16 & 0.441 & 0.002 & 3.099 & 0.388 & -1.337&  0.125 & 0.513773 & 0.0000 & 0.0000 &* \\
17 & 0.542 & 0.002 & 2.620 & 0.373 & -0.173&  0.142 & 0.474797 & 0.0000 & 0.0000 &* \\
18 & 0.799 & 0.002 & 2.567 & 0.359 & -1.020&  0.140 & 0.439750 & 0.0000 & 0.0000 &* \\
19 & 0.270 & 0.003 & 2.031 & 0.345 & -0.856&  0.170 & 0.406977 & 0.0001 & 0.0731 &* \\
20 & 0.607 & 0.002 & 2.406 & 0.333 & 0.687 & 0.138 & 0.378262 & 0.0000 & 0.0002 & \\
21 & 0.572 & 0.002 & 2.524 & 0.321 & 1.521 & 0.127 & 0.353378 & 0.0000 & 0.0001 & \\
22 & 1.212 & 0.002 & 2.042 & 0.309 & -1.481&  0.151 & 0.327324 & 0.0000 & 0.0000 &* \\
23 & 1.185 & 0.003 & 1.904 & 0.299 & 0.573 & 0.157 & 0.304987 & 0.0000 & 0.0000 & \\
24 & 1.056 & 0.002 & 1.864 & 0.289 & 0.991 & 0.155 & 0.285933 & 0.0000 & 0.0000 &* \\
25 & 0.234 & 0.002 & 1.874 & 0.280 & 1.442 & 0.150 & 0.268622 & 0.0000 & 0.2550 & \\
26 & 0.492 & 0.002 & 1.798 & 0.271 & -0.749&  0.151 & 0.251324 & 0.0000 & 0.2587 & \\
27 & 1.000 & 0.003 & 1.616 & 0.262 & -0.643&  0.162 & 0.235295 & 0.0000 & 0.0074 & \\
28 & 1.264 & 0.002 & 1.877 & 0.254 & 0.034 & 0.135 & 0.220528 & 0.0000 & 0.0000 &* \\
29 & 0.112 & 0.003 & 1.550 & 0.246 & 0.643 & 0.158 & 0.206262 & 0.0000 & 0.9622 & \\
30 & 1.334 & 0.002 & 1.786 & 0.237 & 0.060 & 0.133 & 0.191779 & 0.0000 & 0.0000 &* \\
31 & 0.860 & 0.002 & 1.552 & 0.229 & 1.343 & 0.147 & 0.178966 & 0.0000 & 0.1327 & \\
32 & 1.359 & 0.002 & 1.598 & 0.221 & -0.486&  0.139 & 0.167517 & 0.0000 & 0.0000 & \\
33 & 1.558 & 0.003 & 1.362 & 0.213 & -0.844&  0.157 & 0.155760 & 0.0000 & 0.0000 &* \\
34 & 0.039 & 0.002 & 1.383 & 0.207 & 0.563 & 0.150 & 0.146522 & 0.0000 & 0.9999 & \\
35 & 0.181 & 0.003 & 1.258 & 0.200 & -0.344&  0.159 & 0.137192 & 0.0000 & 1.0000 & \\
36 & 0.716 & 0.002 & 1.370 & 0.195 & 0.033 & 0.142 & 0.129616 & 0.0000 & 0.9343 & \\
37 & 1.890 & 0.003 & 1.124 & 0.189 & 1.140 & 0.168 & 0.122645 & 0.0000 & 0.0000 &* \\
38 & 1.686 & 0.003 & 1.079 & 0.184 & 0.051 & 0.171 & 0.116239 & 0.0001 & 0.0000 & \\
39 & 0.951 & 0.003 & 1.002 & 0.180 & 0.718 & 0.180 & 0.110602 & 0.0003 & 1.0000 & \\
40 & 0.513 & 0.003 & 0.962 & 0.175 & 1.345 & 0.182 & 0.105189 & 0.0005 & 1.0000 & \\
41 & 0.754 & 0.003 & 1.032 & 0.171 & -1.246&  0.166 & 0.100534 & 0.0000 & 1.0000 & \\
42 & 0.835 & 0.003 & 0.970 & 0.167 & 0.723 & 0.172 & 0.095679 & 0.0001 & 1.0000 & \\
43 & 1.299 & 0.003 & 0.865 & 0.163 & 0.868 & 0.188 & 0.090910 & 0.0011 & 0.8833 & \\
44 & 0.158 & 0.003 & 0.848 & 0.160 & -1.205&  0.188 & 0.087160 & 0.0011 & 1.0000 & \\
45 & 1.151 & 0.003 & 0.844 & 0.156 & -0.048&  0.185 & 0.083626 & 0.0007 & 1.0000 & \\
46 & 1.865 & 0.003 & 0.860 & 0.153 & 0.208 & 0.178 & 0.080134 & 0.0002 & 0.0002 & \\
47 & 2.085 & 0.003 & 0.834 & 0.150 & 0.148 & 0.180 & 0.076910 & 0.0003 & 0.0000 &* \\
48 & 0.342 & 0.003 & 0.819 & 0.147 & 0.994 & 0.179 & 0.073469 & 0.0003 & 1.0000 & \\
49 & 1.617 & 0.003 & 0.797 & 0.143 & -1.443&  0.180 & 0.070242 & 0.0003 & 0.1162 & \\
50 & 0.626 & 0.003 & 0.785 & 0.140 & -0.741&  0.178 & 0.067099 & 0.0002 & 1.0000 & \\
51 & 0.680 & 0.003 & 0.771 & 0.137 & -0.905&  0.177 & 0.064008 & 0.0002 & 1.0000 & \\
52 & 0.089 & 0.003 & 0.694 & 0.134 & -0.199&  0.193 & 0.061305 & 0.0020 & 1.0000 & \\
53 & 1.781 & 0.003 & 0.679 & 0.131 & -1.315&  0.193 & 0.058821 & 0.0021 & 0.1551 & \\
54 & 1.412 & 0.003 & 0.656 & 0.128 & 1.270 & 0.196 & 0.056409 & 0.0029 & 1.0000 &* \\
55 & 0.303 & 0.003 & 0.690 & 0.126 & -1.245&  0.182 & 0.054027 & 0.0004 & 1.0000 & \\
56 & 0.922 & 0.003 & 0.649 & 0.123 & -0.369&  0.189 & 0.051694 & 0.0013 & 1.0000 & \\
57 & 1.948 & 0.003 & 0.650 & 0.120 & 0.673 & 0.185 & 0.049628 & 0.0007 & 0.0324 &* \\
58 & 1.524 & 0.003 & 0.641 & 0.118 & -0.191&  0.184 & 0.047574 & 0.0006 & 0.9958 & \\
59 & 0.460 & 0.003 & 0.735 & 0.115 & -0.261&  0.157 & 0.045514 & 0.0000 & 1.0000 & \\
60 & 0.418 & 0.003 & 0.634 & 0.113 & 0.752 & 0.178 & 0.043567 & 0.0002 & 1.0000 & \\
61 & 2.044 & 0.003 & 0.585 & 0.110 & 0.785 & 0.188 & 0.041588 & 0.0011 & 0.0555 & \\
62 & 1.232 & 0.003 & 0.561 & 0.108 & 0.050 & 0.192 & 0.039811 & 0.0019 & 1.0000 & \\
63 & 2.312 & 0.003 & 0.576 & 0.106 & -0.542&  0.183 & 0.038170 & 0.0005 & 0.0011 &* \\
64 & 1.086 & 0.003 & 0.612 & 0.103 & 0.959 & 0.169 & 0.036539 & 0.0000 & 1.0000 & \\
65 & 1.022 & 0.003 & 0.585 & 0.101 & -0.733&  0.173 & 0.034927 & 0.0001 & 1.0000 & \\
66 & 0.653 & 0.003 & 0.520 & 0.099 & 0.338 & 0.190 & 0.033229 & 0.0013 & 1.0000 & \\
67 & 1.587 & 0.003 & 0.521 & 0.097 & 0.526 & 0.185 & 0.031866 & 0.0007 & 1.0000 & \\
68 & 2.825 & 0.003 & 0.510 & 0.095 & -1.458&  0.185 & 0.030561 & 0.0007 & 0.0000 &* \\
69 & 2.446 & 0.003 & 0.496 & 0.093 & -0.521&  0.186 & 0.029301 & 0.0008 & 0.0045 & \\
70 & 0.210 & 0.003 & 0.479 & 0.091 & 0.169 & 0.189 & 0.028069 & 0.0012 & 1.0000 & \\
71 & 2.503 & 0.003 & 0.463 & 0.089 & -0.851&  0.192 & 0.026939 & 0.0017 & 0.0092 & \\
72 & 0.979 & 0.003 & 0.431 & 0.087 & 1.115 & 0.202 & 0.025891 & 0.0061 & 1.0000 & \\
73 & 2.011 & 0.003 & 0.431 & 0.085 & -1.410&  0.198 & 0.024960 & 0.0039 & 0.9923 & \\
74 & 3.314 & 0.003 & 0.425 & 0.084 & -0.700&  0.197 & 0.024034 & 0.0035 & 0.0000 & \\
75 & 1.816 & 0.003 & 0.399 & 0.082 & 1.367 & 0.206 & 0.023132 & 0.0094 & 1.0000 & \\
76 & 3.061 & 0.003 & 0.386 & 0.081 & 0.343 & 0.209 & 0.022321 & 0.0135 & 0.0002 & \\
77 & 0.599 & 0.003 & 0.390 & 0.079 & 0.078 & 0.203 & 0.021571 & 0.0072 & 1.0000 & \\
78 & 1.488 & 0.003 & 0.366 & 0.078 & 1.451 & 0.213 & 0.020854 & 0.0193 & 1.0000 & \\
79 & 1.120 & 0.003 & 0.382 & 0.077 & -0.650&  0.201 & 0.020170 & 0.0055 & 1.0000 & \\
80 & 2.274 & 0.003 & 0.367 & 0.075 & -0.571&  0.205 & 0.019467 & 0.0089 & 0.9689 & \\
81 & 1.669 & 0.003 & 0.357 & 0.074 & -0.232&  0.207 & 0.018776 & 0.0110 & 1.0000 & \\
82 & 1.453 & 0.003 & 0.354 & 0.073 & -1.278&  0.206 & 0.018191 & 0.0097 & 1.0000 & \\
83 & 2.136 & 0.003 & 0.336 & 0.072 & 0.383 & 0.214 & 0.017585 & 0.0204 & 1.0000 & \\
84 & 0.484 & 0.003 & 0.329 & 0.071 & -0.148&  0.214 & 0.017009 & 0.0212 & 1.0000 & \\
85 & 3.032 & 0.004 & 0.303 & 0.069 & -0.882&  0.229 & 0.016449 & 0.0752 & 0.0782 & \\
86 & 3.219 & 0.004 & 0.303 & 0.068 & -0.424&  0.226 & 0.015969 & 0.0578 & 0.0115 & \\
87 & 0.800 & 0.004 & 0.298 & 0.067 & 0.859 & 0.226 & 0.015523 & 0.0593 & 1.0000 & \\
88 & 2.630 & 0.004 & 0.288 & 0.066 & 0.222 & 0.230 & 0.015091 & 0.0818 & 0.9744 & \\
89 & 2.362 & 0.004 & 0.281 & 0.066 & 0.836 & 0.233 & 0.014691 & 0.0986 & 1.0000 & \\
90 & 1.736 & 0.004 & 0.280 & 0.065 & 0.991 & 0.231 & 0.014304 & 0.0882 & 1.0000 & \\
91 & 0.026 & 0.004 & 0.282 & 0.064 & 0.440 & 0.227 & 0.013914 & 0.0615 & 1.0000 & \\
92 & 3.000 & 0.004 & 0.272 & 0.063 & 1.194 & 0.231 & 0.013503 & 0.0854 & 0.4292 & \\
93 & 3.267 & 0.004 & 0.275 & 0.062 & -0.866&  0.225 & 0.013120 & 0.0550 & 0.0465 & \\
94 & 2.686 & 0.004 & 0.268 & 0.061 & -1.529&  0.228 & 0.012741 & 0.0664 & 0.9950 & \\
95 & 3.969 & 0.004 & 0.256 & 0.060 & -0.320&  0.235 & 0.012373 & 0.1103 & 0.0001 & \\
96 & 1.377 & 0.004 & 0.250 & 0.059 & -0.586 & 0.237 & 0.012041 & 0.1328 & 1.0000 & \\
97 & 2.214 & 0.004 & 0.242 & 0.059 & 1.496 & 0.242 & 0.011721 & 0.1746 & 1.0000 & \\
98 & 0.178 & 0.004 & 0.237 & 0.058 & -0.223&  0.243 & 0.011419 & 0.1929 & 1.0000 & \\
99 & 2.734 & 0.004 & 0.249 & 0.057 & 1.000 & 0.229 & 0.011133 & 0.0755 & 0.9996 & \\
100 & 0.710 & 0.004 & 0.245 & 0.056 & -1.398&  0.230 & 0.010865 & 0.0785 & 1.0000 & \\
101 & 0.896 & 0.004 & 0.246 & 0.056 & 1.394 & 0.226 & 0.010589 & 0.0589 & 1.0000 & \\
102 & 4.508 & 0.004 & 0.225 & 0.055 & -0.565&  0.244 & 0.010298 & 0.1966 & 0.0000 & \\
103 & 1.175 & 0.004 & 0.224 & 0.054 & 0.888 & 0.242 & 0.010044 & 0.1736 & 1.0000 & \\
104 & 2.571 & 0.004 & 0.223 & 0.054 & 0.064 & 0.240 & 0.009795 & 0.1582 & 1.0000 & \\
105 & 2.172 & 0.004 & 0.205 & 0.053 & -0.714&  0.257 & 0.009551 & 0.3888 & 1.0000 & \\
106 & 4.709 & 0.004 & 0.204 & 0.052 & -0.305&  0.256 & 0.009339 & 0.3664 & 0.0000 & \\
107 & 2.758 & 0.004 & 0.201 & 0.052 & 1.071 & 0.257 & 0.009131 & 0.3820 & 1.0000 & \\
108 & 3.675 & 0.004 & 0.198 & 0.051 & 0.702 & 0.258 & 0.008926 & 0.3961 & 0.3397 & \\
109 & 4.092 & 0.004 & 0.201 & 0.051 & 0.846 & 0.252 & 0.008729 & 0.3006 & 0.0137 & \\
110 & 3.817 & 0.004 & 0.199 & 0.050 & 1.131 & 0.251 & 0.008536 & 0.2862 & 0.1327 & \\
111 & 2.417 & 0.004 & 0.192 & 0.049 & -0.180&  0.258 & 0.008340 & 0.3912 & 1.0000 & \\
112 & 3.174 & 0.004 & 0.190 & 0.049 & -0.642&  0.257 & 0.008154 & 0.3775 & 0.9999 & \\
113 & 1.698 & 0.004 & 0.188 & 0.048 & -0.220&  0.258 & 0.007975 & 0.3893 & 1.0000 & \\
114 & 0.544 & 0.004 & 0.185 & 0.048 & -1.488&  0.258 & 0.007796 & 0.4052 & 1.0000 & \\
115 & 4.881 & 0.004 & 0.182 & 0.047 & -1.268&  0.259 & 0.007625 & 0.4108 & 0.0001 & \\
116 & 1.266 & 0.004 & 0.176 & 0.047 & -0.486&  0.265 & 0.007458 & 0.5180 & 1.0000 & \\
117 & 4.338 & 0.004 & 0.176 & 0.046 & 1.237 & 0.262 & 0.007297 & 0.4611 & 0.0252 & \\
118 & 0.399 & 0.004 & 0.169 & 0.046 & -1.036&  0.270 & 0.007136 & 0.6133 & 1.0000 & \\
119 & 3.413 & 0.004 & 0.173 & 0.045 & 1.460 & 0.262 & 0.006993 & 0.4597 & 0.9995 & \\
120 & 2.063 & 0.004 & 0.169 & 0.045 & 0.953 & 0.265 & 0.006849 & 0.5155 & 1.0000 & \\
121 & 4.440 & 0.004 & 0.169 & 0.044 & -0.939&  0.262 & 0.006704 & 0.4711 & 0.0279 & \\
122 & 1.628 & 0.004 & 0.160 & 0.044 & -0.387&  0.274 & 0.006564 & 0.6661 & 1.0000 & \\
123 & 2.938 & 0.004 & 0.175 & 0.043 & -0.283&  0.248 & 0.006436 & 0.2508 & 1.0000 & \\
124 & 2.600 & 0.004 & 0.163 & 0.043 & -0.035&  0.264 & 0.006313 & 0.4980 & 1.0000 & \\
125 & 3.134 & 0.004 & 0.159 & 0.043 & 1.172 & 0.267 & 0.006186 & 0.5466 & 1.0000 & \\
126 & 0.756 & 0.004 & 0.157 & 0.042 & 0.189 & 0.267 & 0.006061 & 0.5585 & 1.0000 & \\
127 & 2.802 & 0.004 & 0.156 & 0.042 & 0.559 & 0.268 & 0.005942 & 0.5613 & 1.0000 & \\
128 & 3.732 & 0.004 & 0.150 & 0.041 & 1.278 & 0.274 & 0.005821 & 0.6765 & 0.9990 & \\
129 & 0.358 & 0.004 & 0.149 & 0.041 & 0.198 & 0.274 & 0.005706 & 0.6692 & 1.0000 & \\
130 & 1.326 & 0.004 & 0.149 & 0.040 & -0.873&  0.272 & 0.005597 & 0.6370 & 1.0000 & \\
131 & 1.059 & 0.004 & 0.145 & 0.040 & 0.175 & 0.276 & 0.005486 & 0.7007 & 1.0000 & \\
132 & 1.857 & 0.003 & 0.190 & 0.040 & 1.567 & 0.209 & 0.005379 & 0.0126 & 1.0000 & \\
133 & 1.902 & 0.003 & 0.193 & 0.039 & -1.303&  0.204 & 0.005273 & 0.0075 & 1.0000 & \\
134 & 1.975 & 0.004 & 0.161 & 0.039 & -1.506&  0.241 & 0.005137 & 0.1689 & 1.0000 & \\
135 & 0.141 & 0.004 & 0.147 & 0.038 & 1.142 & 0.260 & 0.005012 & 0.4364 & 1.0000 & \\
136 & 3.601 & 0.004 & 0.140 & 0.038 & -0.660&  0.270 & 0.004907 & 0.6029 & 1.0000 & \\
137 & 6.593 & 0.004 & 0.140 & 0.037 & -0.019&  0.268 & 0.004809 & 0.5695 & 1.0000 & \\
138 & 4.216 & 0.004 & 0.137 & 0.037 & 0.991 & 0.271 & 0.004711 & 0.6196 & 0.9020 & \\
139 & 3.538 & 0.004 & 0.142 & 0.037 & -0.623&  0.259 & 0.004618 & 0.4214 & 1.0000 & \\
140 & 8.228 & 0.004 & 0.134 & 0.036 & -1.086&  0.272 & 0.004529 & 0.6323 & 1.0000 & \\
141 & 0.844 & 0.004 & 0.130 & 0.036 & 0.465 & 0.277 & 0.004440 & 0.7262 & 1.0000 & \\
142 & 7.411 & 0.004 & 0.132 & 0.036 & 1.112 & 0.271 & 0.004353 & 0.6157 & 1.0000 & \\
143 & 2.475 & 0.004 & 0.129 & 0.035 & -1.353&  0.273 & 0.004267 & 0.6604 & 1.0000 & \\
144 & 3.774 & 0.004 & 0.133 & 0.035 & 1.255 & 0.262 & 0.004187 & 0.4734 & 1.0000 & \\
145 & 3.927 & 0.004 & 0.129 & 0.035 & 0.480 & 0.269 & 0.004108 & 0.5867 & 1.0000 & \\
146 & 3.370 & 0.004 & 0.126 & 0.034 & 0.044 & 0.273 & 0.004027 & 0.6588 & 1.0000 & \\
147 & 3.446 & 0.004 & 0.134 & 0.034 & 0.620 & 0.254 & 0.003948 & 0.3289 & 1.0000 & \\
148 & 5.646 & 0.004 & 0.133 & 0.034 & 0.178 & 0.252 & 0.003858 & 0.2980 & 0.0005 & \\
149 & 1.023 & 0.004 & 0.121 & 0.033 & -0.332&  0.276 & 0.003782 & 0.7016 & 1.0000 & \\
150 & 5.698 & 0.004 & 0.127 & 0.033 & 0.742 & 0.259 & 0.003708 & 0.4120 & 0.0012 & \\
\end{longtable}
\end{landscape}}

\end{appendix}

\end{document}